\pdfoutput=1

\documentclass[11pt]{article}
\usepackage{acl}

\usepackage{times}
\usepackage{latexsym}
\usepackage[T1]{fontenc}
\usepackage[utf8]{inputenc}
\usepackage{microtype}
\usepackage{inconsolata}
\usepackage{amsmath,amssymb}
\usepackage{booktabs}
\usepackage{graphicx}
\usepackage{multirow}
\usepackage[dvipsnames]{xcolor}
\usepackage{url}
\usepackage{wrapfig}
\usepackage{float}
\usepackage{array}
\usepackage{caption}
\usepackage{xspace}
\usepackage{algorithm}
\usepackage{algpseudocode}
\usepackage{hyperref}
\usepackage[nameinlink,noabbrev]{cleveref}

\definecolor{darkblue}{rgb}{0, 0, 0.5}
\hypersetup{colorlinks=true, citecolor=darkblue, linkcolor=darkblue, urlcolor=darkblue}


\usepackage{tcolorbox}
\tcbuselibrary{listings,skins,breakable}

\newtcblisting{promptbox}[1][]{
    enhanced,
    breakable,
    attach boxed title to top center={yshift=-3mm,yshifttext=-1mm},
    colback=white,
    colframe=black,
    colbacktitle=black!75!yellow,
    boxed title style={size=small},
    listing only,
    listing options={
        breaklines=true,        
        breakatwhitespace=true, 
        basicstyle=\small\ttfamily,
        breakindent=0pt
    },
    #1
}

\lstdefinestyle{boldwords}{
  basicstyle=\scriptsize\ttfamily,
  moredelim=[is][\bfseries]{**}{**} 
}

\newtcblisting{mainpromptbox}[1][]{
    enhanced,
    width=0.49\textwidth,
    attach boxed title to top center={yshift=-3mm,yshifttext=-1mm},
    colback=white,
    colframe=black,
    colbacktitle=black!75!yellow,
    boxed title style={size=small},
    listing only,
    listing options={
        breaklines=true,
        breakatwhitespace=true,
        breakindent=0pt,
        style=boldwords
    },
    left=5pt,          
    right=0pt,         
    top=0pt,           
    bottom=0pt,        
    #1
}

\colorlet{colexam}{brown!50!black}
\tcbset{
  base/.style={
    empty,
    frame engine=path,
    colframe=brown!10,
    sharp corners,
    title={Generated Rationale},
    attach boxed title to top left={yshift*=-\tcboxedtitleheight},
    boxed title style={size=minimal, top=4pt, left=4pt},
    coltitle=colexam,
    fonttitle=\small\bfseries\sffamily,
  }
}
\newtcolorbox[use counter=example]{myexamplea}{%
  base,
  colback=colexam,
  boxed title style={
    overlay={
      \draw[colexam,line width=2pt] (frame.north west)--(frame.north east);
    }
  },
  overlay unbroken={
    \draw[colexam] ([yshift=-1.5pt]title.north east)--([xshift=-0.5pt, yshift=-1.5pt]title.north-|frame.east);
  }
}

\title{\textsc{Graft}: Graph-Distilled Generative Retrieval for Facet-Aware Scientific Literature Exploration}
\author{Italo Luis da Silva, Hanqi Yan, Yujing Wang, Jiangnan Ye, Lin Gui \& Yulan He \\
Department of Informatics\\
King's College London\\
London, England, UK \\
\texttt{\{italo.da\_silva,hanqi.yan,jiangnan.ye,lin.1.gui,yulan.he\}@kcl.ac.uk} \\
\texttt{eugeniawyj@outlook.com}
}

\newcommand{\Dataset}{\textsc{LitWeave}\xspace}
\newcommand{\Method}{\textsc{Graft}\xspace}
\newcommand{\facets}{\mathcal{F}}

\begin{document}

\maketitle
\begin{abstract}

Scientific papers may relate by problem, method, result, or contribution, but
    document-level retrievers collapse these into a single similarity score without
    saying why they are related. Citation- and similarity-based retrieval alone also
    confines search to the neighbourhood of what is already known, whereas generative
    retrieval generates document identifiers directly, enabling the exploratory
    retrieval that scientific discovery depends on. We connect papers in a graph whose
    edges are typed by these four facets, derived from facet items and citation signals,
    and distil it into a generative retriever whose identifiers are the papers' own
    facet text. Two graph properties do not survive naive distillation. First, because
    every training pair is an edge, naive enumeration indexes just 84\% of the corpus.
    \emph{Coverage-aware distillation} makes every paper learnable through a
    reverse-neighbour fallback, a minimum-coverage threshold and edge-importance
    weighting. Second, constrained decoding guarantees that every generated identifier
    is a valid paper, but not that the graph connects it to the query.
    \emph{Graph-weighted reciprocal rank fusion} scales each candidate's rank term by
    its query--candidate edge weight, dropping unsupported ones. On \Dataset, our
    constructed corpus of 11,359 NLP papers, \Method recovers 91\% of its graph
    teacher's Recall@20 with no nearest-neighbour index or encoder at inference, and
    outperforms the graph teacher on query papers outside the corpus. It reproduces the
    graph's own facet labels at 0.922 precision, so every returned paper arrives
    labelled with the facet that surfaced it rather than an opaque score.

\end{abstract}

\section{Introduction}
\label{sec:introduction}

Large Language Models (LLMs) have increasingly been used in scientific research across
many different parts of the process, such as novelty
assessment~\citep{lin2024evaluatingenhancinglargelanguage,shahid-etal-2025-literature,wu2026novbench,Silva2026GraphMind},
idea
generation~\citep{si2025can,wang2024scimonscientificinspirationmachines,gu2025interestingscientificideageneration},
automated literature
reviews~\citep{wang2024autosurvey,liang2025surveyx,yan2025surveyforge}, peer-review
writing~\citep{jin2024agentreviewexploringpeerreview,chitale2025autorev} and even
autonomous research and
communication~\citep{Lu2024TheAS,ghafarollahi2024sciagentsautomatingscientificdiscovery}.
A common requirement across these tasks is high-quality retrieval of relevant scientific
papers from the existing literature.

\begin{figure*}[t]
\centering
\includegraphics[width=\textwidth]{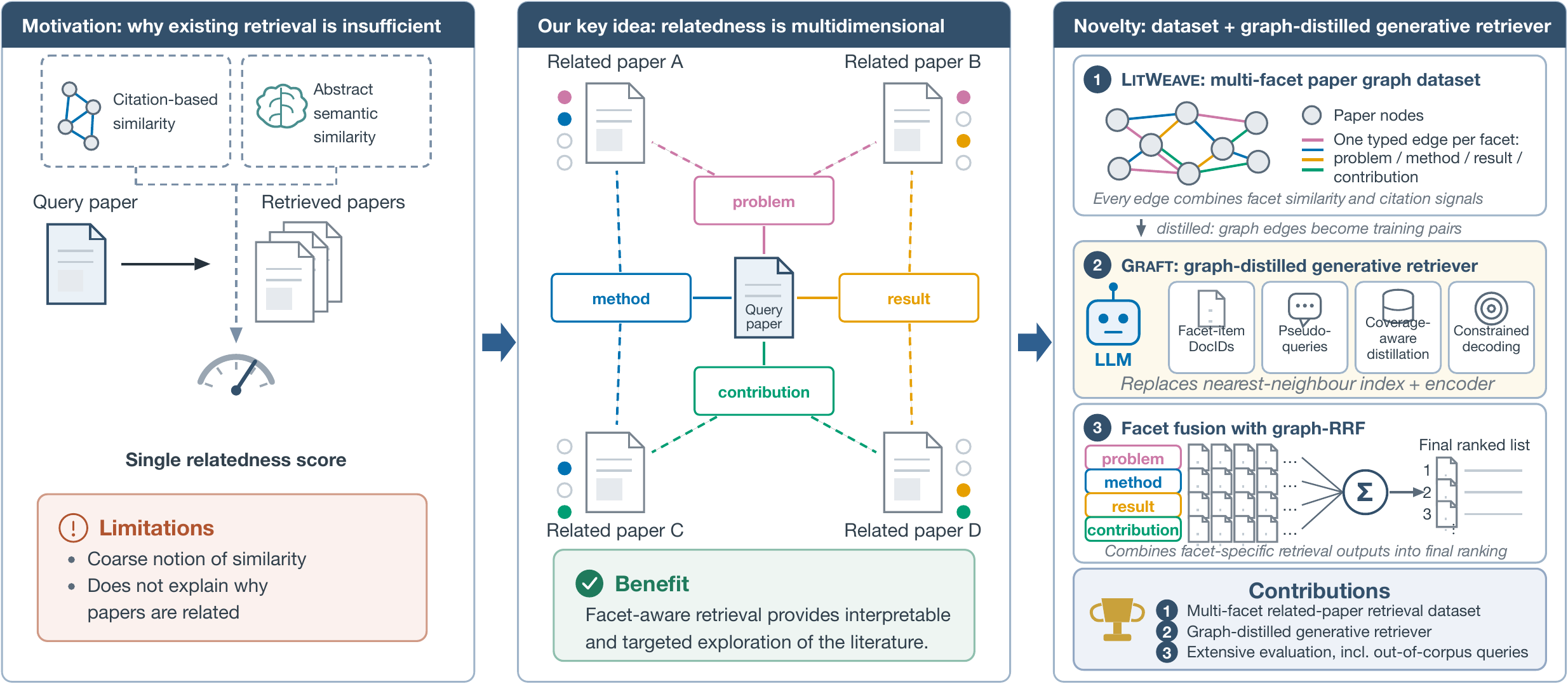}
\caption{\textbf{Left:} existing retrieval reduces relatedness to a single score and
    cannot say why two papers are related. \textbf{Centre:} \Dataset{} instead types
    every edge by facet, so a pair may align on one facet or on several.
    \textbf{Right:} \Method{} distils that graph into a generative retriever and fuses
    the per-facet lists with graph-RRF, so the final ranking keeps the typing. The
    diagram is schematic; measured results are in \S\ref{sec:experiments}.}
\label{fig:overview}
\end{figure*}

While standard retrieval systems return passages closely related to a given
query~\citep{dpr-karpukhin-2020}, in scientific related-paper retrieval the query is itself a paper and relatedness is
multi-dimensional: two papers may address the same problem while adopting entirely
different methods, or share an evaluation protocol despite targeting different research
questions. Consider a benchmark for evaluating counterfactual text
generation~\citep{ceval-nguyen-2024}: some related papers are other
counterfactual-generation work, but another is an LLM-as-a-judge
framework~\citep{yescieval-dsouza-2025} with no counterfactual content at all, related
because of \emph{how} the benchmark evaluates, not \emph{what} it evaluates. Existing
retrieval systems typically collapse these heterogeneous relationships into a single
similarity score~\citep{specter-cohan-2020}, making it difficult to distinguish why a paper is relevant. This
motivates facet-aware retrieval, where relatedness is modelled explicitly through
semantic facets such as problems, methods, results, and contributions
(Figure~\ref{fig:overview}).

Therefore, in this paper, we tackle two research questions toward building such a
system: (1) how should these heterogeneous relationships be represented? and (2) how can
such structured knowledge be efficiently indexed for retrieval?

To answer the first question, we represent the literature as a typed paper graph, where
papers are connected through facet-specific edges corresponding to problems, methods,
results, and contributions. This graph preserves both the type and strength of each
relationship, providing an explicit structural representation of facet-aware
relatedness. To support this formulation, we construct \Dataset, a graph-structured
scientific retrieval corpus that combines semantic facet extraction with citation
signals.

The second question arises because representation alone is not enough: the graph must be
indexed so that its structured knowledge can be efficiently accessed during retrieval.
While retrieval directly over the graph can exploit this structure, it still relies on
dense encoders and nearest-neighbour search at inference. We instead propose \Method,
which distills the graph into a generative retriever, allowing the model itself to
become the index~\citep{dsi-tay-2022} while preserving facet-aware retrieval behaviour.
Naively distilling the graph, however, fails to preserve two key structural properties.
First, graph sparsity leads to incomplete supervision, leaving many papers
underrepresented during training; generative retrieval typically closes such gaps with
self-referential targets~\citep{bridging-zhuang-2023}, but our queries are themselves
papers in the graph, so a self-target would just teach the model to echo the query's own
identifier. Second, although constrained generation ensures valid document
identifiers~\citep{constrained-wu-2025}, it cannot guarantee that generated papers remain
structurally supported by the original graph. We therefore introduce coverage-aware distillation to preserve graph coverage and
graph-weighted reciprocal rank fusion (graph-RRF) to enforce graph-supported retrieval
during inference.

To explore this formulation, we instantiate \Dataset with 11,359 NLP papers from the
Semantic Scholar Open Research Corpus~\citep{s2orc-lo-2020}, connecting papers both by
citation (co-citation and bibliographic coupling) and semantic relationships defined
over LLM-extracted facets. We evaluate \Method on \Dataset against lexical, dense,
graph-based and generative baselines: it reaches 91\% of the dense-graph teacher's
Recall@20 and outperforms every baseline that does not consult the graph. On query
papers outside the corpus, evaluated against their bibliographies, this ordering
reverses: \Method leads the teacher and every other baseline at every cutoff, by 0.149
Recall@5, while needing only the query's facet text where the teacher must score the
query against its whole index. Coverage-aware distillation contributes 0.030 R@20
(+10.13\%) over uniform sampling, two thirds of it from the coverage mechanisms rather
than from edge weighting, and graph-RRF adds 0.038 R@20 (+13.38\%) over plain fusion
while reproducing the graph's facet attribution at 0.922 precision. Ablations further
cover identifier design, objective functions, and training strategies.

In summary, our contributions are:
\begin{enumerate}

    \setlength{\itemsep}{0pt}
    \setlength{\parskip}{0pt}
    \setlength{\parsep}{0pt}

    \item \Dataset, a multi-facet corpus for scientific related-paper retrieval: 11,359
        NLP papers connected as a typed graph over four semantic facets, where every
        edge indicates which facet relates two papers and how strongly.

    \item \Method, a framework for distilling that graph into a generative retriever,
        with \emph{coverage-aware distillation} and \emph{graph-weighted RRF} as its two
        components addressing the coverage and grounding gaps above.

    \item An empirical study comparing \Method against lexical, dense, graph-based and
        generative baselines, with ablations of identifier design, objective functions
        and training strategies, including an out-of-corpus evaluation, where the
        distilled retriever overtakes the teacher on unseen papers.

\end{enumerate}

\section{Related Work}
\label{sec:related-work}

\paragraph{Scientific related-paper retrieval.}
\label{sec:rw-retrieval}

Scientific retrieval has been approached through lexical, dense and citation-based
representations. Lexical methods such as BM25~\citep{bm25-robertson-2009} remain strong
baselines because they preserve exact terminology. Dense methods encode titles and
abstracts as vector embeddings~\citep{reimers-2019-sentence-bert}, and
scientific-document models use citation structure to learn task-specific
representations~\citep{specter-cohan-2020,Singh2022SciRepEvalAM,scincl-ostendorff-2022}.
Recent work makes the retrieval signal more explicit through corpus-grounded scientific
concepts~\citep{zhang-etal-2025-scientific,do2026casperconceptintegrated}. Others employ
aspect-based similarity, comparing papers through facets, such as their methods or
findings~\citep{ostendorff2020aspect,aspire-mysore-2022}. A more established line uses
citation networks, where bibliographic coupling~\citep{kessler-1963} and
co-citation~\citep{small-1973} capture relatedness beyond direct citation links, with
later work refining this approach using full-text citation context and
intent~\citep{kleminski2022citation,zhang2025improvingbibliographic,phan2026citationintentrelatedness}.
These methods return relatedness at the level of whole documents, without indicating the
source of relations. Our corpus instead connects papers by both citation features and
per-facet semantic similarity, so relatedness carries facet-level provenance.

\paragraph{Generative retrieval.}
\label{sec:rw-genret}

Generative retrieval formulates search as generation: instead of encoding a query and
searching an external index, the model generates the identifiers of relevant documents
directly, such that the model becomes the index. DSI~\citep{dsi-tay-2022} introduced
this idea, and later work expanded it through neural corpus indexing, n-gram and
semantic identifiers, and end-to-end
training~\citep{nci-wang-2022,seal-bevilacqua-2022,ultron-zhou-2022,corpusbrain-chen-2022}.
A central design choice is how to identify documents, such as using atomic labels or
semantic identifiers where related items share parts of their
code~\citep{tiger-rajput-2023,zhang2025purelysemanticindexing,zhang2025c2tid,mekonnen2025ddro},
or lexical identifiers learned from document terms~\citep{glen-lee-2023}. Closest to our
method, MINDER~\citep{minder-li-2023} represents a document through several views,
including pseudo-queries. We evaluate MINDER as a baseline in \S\ref{sec:main-results}
and \S\ref{sec:attribution}. DSI-QG~\citep{bridging-zhuang-2023} pairs every document
with pseudo-queries from its own text, so each is trained as a self-target; our pairs
are edges between distinct papers, so coverage instead follows in-degree. Constrained
decoding guarantees well-formed identifiers, not relevant
ones~\citep{constrained-wu-2025}, which is what graph-RRF checks, against a structure
outside the model rather than the model's own confidence.

\section{Task and Dataset}
\label{sec:task-and-dataset}

We address related-paper retrieval with facet attribution: given a query paper $p$ and a
corpus, retrieve the papers most related to $p$ by querying each facet $f \in \facets$
separately and combining the per-facet results. What distinguishes the task is
\textit{provenance}: alongside which papers are most relevant, the result records which
facets retrieved each one.

\subsection{Corpus Construction}
\label{sec:corpus-construction}

\Dataset connects papers by the semantic facets that describe them, giving a typed graph
where each edge carries a facet and an importance weight (\S\ref{sec:graph-teacher}).

We draw papers from the Semantic Scholar Open Research Corpus
(S2ORC)~\citep{s2orc-lo-2020}: \textsc{*ACL} and related NLP venues, published
2019--2026, with parsed full text, of which 27,757 qualify. Given the complexity of
further processing steps, we downsample the dataset. Random sampling would break the
citation structure, since an edge survives only when both endpoints are kept, so we take
a $k$-core and select papers from it under per-year quotas summing to 10,000, then add
2025--26 papers separately, since they are too recent to be retained by either
procedure. This gives 11,359 papers, split temporally into train (2019--2022, 6,875),
dev (2023, 1,543) and test (2024--2026, 2,941). Appendix~\ref{app:td-corpus} describes
the procedure and Table~\ref{tab:app-year-split} the per-year breakdown.

\subsection{Facet Extraction and Citation Signals} \label{sec:facets-and-signals}
\paragraph{Facets.} We represent each paper by four \emph{facets}, extracted from its
parsed full text by an LLM (\texttt{gpt-5-mini}): \textbf{Problems}, the research
questions the paper addresses; \textbf{Methods}, the approaches or techniques it
proposes or uses; \textbf{Results}, its main findings; and \textbf{Contributions}, what
it adds to the field. Each facet is a short list of bullet items (average of 3.7--4.2
per facet), and every paper receives at least one item in each facet. Each facet item is
separately encoded as a vector embedding. An automatic audit for structural
well-formedness, grounding in the source text, redundancy and specificity finds over
99\% of papers clean on every check (Appendix~\ref{app:td-validation}). We also validate
groundedness manually: three annotators rated one item per facet from sampled papers,
280 ratings over 200 items on a 1--5 scale. 96.5\% of items are rated ${\geq}\,4$ and
none 1; on a shared 40-item core, 35 items are unanimously rated grounded, with
chance-corrected agreement of 0.91 (Gwet's $\mathrm{AC}_1$;
Appendix~\ref{app:groundedness}).

\paragraph{Citation signals.} Alongside facet similarity, we calculate two signals from
the citation graph: \emph{bibliographic coupling}, the number of references cited by
both papers, and \emph{co-citation}, the number of papers that cite both. These signals
encode the rich relationship between two papers and their interactions across the
literature. Both are heavy-tailed, so we log-compress and average them into a single
citation score (Appendix~\ref{app:edge-weight-selection}).

\subsection{The Multi-Facet Paper Graph}
\label{sec:graph-teacher}

Facets and citation signals combine into a typed graph over the corpus, where each node
is a paper and edges of different types carry the four facets, so that two papers often
connect through several facets at once, which allows us to track their provenance.

\paragraph{Edge scoring.} Each facet item is encoded separately with
\texttt{all-MiniLM-L12-v2}~\citep{reimers-2019-sentence-bert}. The facet score
$s_f(A, B)$ is a ColBERT-inspired maximum similarity~\citep{khattab2020colbert} over the
two papers' items in facet $f$, symmetrised and normalised for item count
(Appendix~\ref{app:edge-weight-selection}). The edge weight combines it with the
citation score $c(A, B)$ of \S\ref{sec:facets-and-signals}, which acts as a floor: two
papers that co-occur often in the literature retain some edge weight even when their
facet score is low. The two are combined with weights $\alpha=0.75$ and $\beta=0.25$,
selected by a sweep on the development split
(Appendix~\ref{app:edge-weight-selection}). Scores are computed over all pairs of the
11,359 papers, but most are low and uninformative, so we keep only the top-20 edges per
paper in each facet.

\paragraph{Retrieval from the graph.} We use this graph both as a baseline
(\S\ref{sec:experiments-setup}) and as the teacher for our method (\S\ref{sec:method}).
Retrieval runs in parallel over the four facets: from the query paper's node in each
facet, a two-hop breadth-first walk follows the precomputed top-20 edges and accumulates
a path score. We fuse the four resulting lists with reciprocal rank
fusion~\citep{cormack2009reciprocal}, giving a candidate list in which each paper carries
its provenance, the set of facets whose walk surfaced it
(Appendix~\ref{app:graph-teacher}).

\section{Graph-Distilled Generative Retrieval}
\label{sec:method}

\begin{figure*}[t]
\centering
\includegraphics[width=0.85\textwidth]{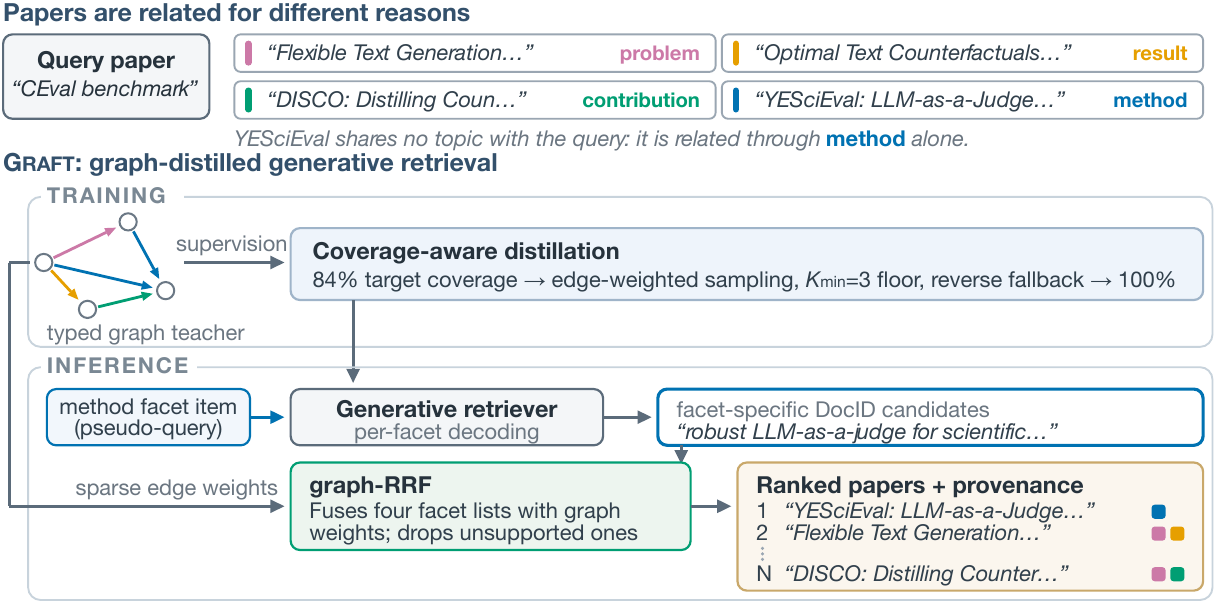}
\caption{\Method on the \emph{CEval} query. \textbf{Top:} different facets connect the
    query to different papers. E.g., a paper on YESciEval is related through
    \emph{method} alone, a connection a document-level dense retriever can miss.
    \textbf{Training:} the typed graph teacher supplies the supervision for
    coverage-aware distillation, which raises target coverage from 84\% to the whole
    corpus. \textbf{Inference:} the same graph supplies the sparse edge weights that
    graph-RRF uses, producing ranked papers whose facet provenance is marked by colour,
    as in the top panel, and which may credit more than one facet. Decoding runs
    separately per facet; the \emph{method} run is shown. The diagram is schematic.}
    \label{fig:methodology}
\end{figure*}

We use the multi-facet paper graph of \S\ref{sec:graph-teacher} as a teacher for our
generative retriever, which decodes facet-specific identifiers for all four facets
(Figure~\ref{fig:methodology}). This raises four key questions: how to represent the
query papers, how to identify the target papers, how to train the retriever, and how to
combine the per-facet results.

\subsection{Natural-Language Facet Identifiers}
\label{sec:natural-language-docids}

The retriever operates on facet items rather than paper text: each one is a
\emph{pseudo-query}, and a paper is represented by one pseudo-query per item. The target
document identifiers (DocIDs) are the same items, normalised and truncated to their
first 12 tokens. Following MINDER~\citep{minder-li-2023}, which represents a document
through several views, we index each paper once per facet item, so a paper is reachable
through several DocIDs, giving the retriever multiple chances to surface it. DocIDs do
not need to be unique: when two papers share a facet item, both are kept and included at
decoding time. Such collisions are rare, and Appendix~\ref{app:identifier-construction}
gives the construction and collision statistics.

At inference, the model receives a prompt naming the facet and one pseudo-query
(Appendix~\ref{app:prompt-format}), and decodes under a prefix-trie constraint that
admits only tokens continuing towards a valid DocID. We decode with beam search and keep
every beam as a candidate; because a paper is reachable through several of its facet
items, we collapse the results into unique papers. Appendix~\ref{app:decoding} gives the
beam settings and how multi-paper identifiers are resolved.

\subsection{Coverage-Aware Distillation}
\label{sec:graph-supervision}

We convert each source paper $q$ and its neighbours $d$ in each facet $f$ into training
tuples $(q, d, f)$: for each query facet item $q_{f, i}$ we sample a target facet item
$d_{f, j}$ and form the pair $(q_{f, i}, d_{f, j})$. Sampling uniformly over
source--neighbour pairs would ignore the edge score, as both strong and weak neighbours
would be seen equally often. Loss weighting is the usual fix, but its improvement is
modest here (\S\ref{sec:coverage-results}). Instead, we follow cost-proportionate
sampling~\citep{zadrozny2003}, oversampling neighbours by 1--5 times, growing linearly
with the edge score relative to the largest edge weight in that facet.
However, oversampling starves the less important neighbours, and a rarely seen DocID is
never generated during retrieval. We therefore impose a minimum target-coverage floor
$K_{\text{min}}$: a DocID seen fewer than $K_{\text{min}}$ times is boosted to that many
samples, with the additional source papers drawn from its gold in-neighbours.

Papers with no incoming edges are never targets at all, and the floor cannot reach them
because it has no in-neighbours to draw from. This is a major issue: only 84\% of the
corpus has an incoming edge in some facet. We recover the rest by flipping their
outgoing edges: given $A$ with no incoming edges but outgoing $A \rightarrow B$, we draw
training examples from $B \rightarrow A$ as a reverse-neighbour fallback. The flip is
viable because $w_f$ is symmetric in value. The graph stores only each paper's top-$K$
neighbours, so an edge present only as $A \rightarrow B$ is a genuine relation that
pruning dropped in the reverse direction. We take the $N$ strongest edges and repeat the
previous steps with them (importance sampling and $K_{\text{min}}$ floor). How much
coverage is missing in the first place depends on the pruning level
(Appendix~\ref{app:pruning}).

Full coverage could also be obtained by training each paper on its own text, as in DSI's
indexing task~\citep{dsi-tay-2022}. This is unsuitable here because the query is itself
a paper: the retriever would learn to return the query's own identifier rather than
those of its neighbours. Both mechanisms therefore use only existing edges: the floor
samples from a target's gold in-neighbours, while the fallback uses its outgoing edges.

This sequence of edge score--proportionate sampling, $K_{\text{min}}$ floor and
reverse-neighbour fallback creates our final training set. We then train the generative
retrieval model with standard cross-entropy on the resulting examples, learning the
distribution of target papers given a facet and a facet-item pseudo-query. In our
setting ($K_{\text{min}} = 3$, $N = 10$), this yields 294,089 training examples, against
213,447 under uniform sampling.

\subsection{Facet Fusion with Graph-RRF}
\label{sec:graph-rrf}

After using the generative retriever to get the candidate lists for each facet, we use
our \emph{graph-RRF} method to combine the different lists into one retrieval result. It
is an extension of reciprocal rank fusion (RRF)~\citep{cormack2009reciprocal}, but while
RRF only takes into account the rank of each item in its original list, we have
another source of information: the paper graph.

This requires only one lightweight sparse adjacency matrix per facet, holding the pruned
top-20 neighbours of each paper (\S\ref{sec:graph-teacher}), from which we read the edge
weight between the query paper and each retrieved candidate. This allows us to fuse the
candidate lists while preserving the valuable information from the graph, prioritising
graph-corroborated candidates. Another consequence is the filtering of candidates that
are not neighbours of the query paper in the graph. The generative model can produce
DocIDs of any paper in the dataset, but not all generated DocIDs are actual neighbours
of the query for that facet, making them low-quality candidates. By using the graph
weights, these candidates have their score set to zero and are filtered out of the top
results.

For a given query $q$, the generative model decodes a ranked candidate list $\mathcal{L}_f(q)$
for each facet $f \in \facets$. We fuse the per-facet lists, assigning each candidate a score according to the following definition:

\begin{equation}
    \mathrm{score}(q,d) = \sum_{f \in \facets \,:\, d \in \mathcal{L}_f(q)}
        \frac{w_f(q,d)}{k + \mathrm{rank}_f(q,d)}
    \label{eq:graph-rrf}
\end{equation}

\noindent where $\mathrm{rank}_f(q,d)$ is the rank of candidate paper $d$ in
$\mathcal{L}_f(q)$ (lower is better), $w_f(q,d)$ is the query--candidate edge weight in
facet $f$'s graph (\S\ref{sec:facets-and-signals}), 0 when no edge exists, and $k=60$
is the standard RRF constant. Each facet contributes a reciprocal-rank term as in plain
RRF, but scaled by the graph weight: a candidate the model ranks highly \emph{and} the
graph corroborates accumulates a large score, whereas a facet on which the graph does
not support the pair contributes nothing.

When querying papers in the corpus, Equation~\ref{eq:graph-rrf} gets $w_f(q,d)$ from the
query's row of the adjacency matrix. The algorithm itself generalises to other settings:
it can use any score expressing how well-supported a candidate $d$ is by a facet $f$.
\S\ref{sec:out-of-corpus} uses this to query out-of-corpus papers by approximating
the weight from the retrieved candidates' neighbours.

\section{Experiments}
\label{sec:experiments}

We organise this section around four questions: how closely \Method matches the graph it
was distilled from, and how it compares with baselines (\S\ref{sec:main-results}); how
well it performs when querying unseen papers (\S\ref{sec:out-of-corpus}); how much
coverage-aware distillation gains over uniform sampling (\S\ref{sec:coverage-results});
and whether graph-RRF preserves the graph's attribution along with its results
(\S\ref{sec:attribution}). \S\ref{sec:further-ablations} considers the impact of
identifier design and model capacity.

\subsection{Setup}
\label{sec:experiments-setup}

\paragraph{Data.}

All experiments use the \Dataset corpus of \S\ref{sec:corpus-construction}, with the
temporal train/dev/test split of Appendix~\ref{tab:app-year-split}. The test split
(2,941 papers) forms the query set. Supervision is held out on the query side: every
training tuple of \S\ref{sec:graph-supervision} is sourced from a training paper, so no
test paper's facet items are ever used as a training query and none of its gold edges
are seen as query-side supervision. Target DocIDs range over the whole corpus: a
training query may point to any paper, test papers included, so the model learns to emit
every DocID, just as a dense retriever indexes every document. The decoding trie covers
the DocIDs of the entire corpus (Appendix~\ref{app:identifier-construction}), and the
baselines retrieve over the same corpus. Gold for each query is its neighbour set in the
full paper graph (\S\ref{sec:facets-and-signals}), and recall measures how much of that
set a method surfaces.

\paragraph{Metrics.}

We report recall@20 and recall@50 (\S\ref{sec:graph-rrf}),
micro-averaged over the held-out queries. Unless stated otherwise, generative retrieval
numbers use graph-RRF.

\paragraph{Baselines.}

We compare against seven systems, all evaluated over the same corpus and gold.
\emph{Lexical}: (i)~\emph{BM25}~\citep{bm25-robertson-2009} over title and facets.
\emph{Document-level dense}: (ii)~\emph{MiniLM}, \texttt{all-MiniLM-L12-v2} embeddings
of title and abstract, and (iii)~\emph{SPECTER2}~\citep{Singh2022SciRepEvalAM}, a
scientific document encoder, also over title and abstract. \emph{Facet-level dense}:
(iv)~\emph{dense-only}, facet embeddings fused without any graph signal, which isolates
the contribution of the graph. \emph{Graph}: (v)~\emph{dense-graph}, the graph retriever
of \S\ref{sec:graph-teacher}. \emph{External}: (vi)~the \emph{Semantic Scholar
Recommendations API}~\citep{s2-recommendations-api}, a production paper-recommendation
service. \emph{Generative}: (vii)~\emph{MINDER}~\citep{minder-li-2023}, with its
multi-view identifiers and language-model-score fusion
(Appendix~\ref{app:minder-baseline}).

\paragraph{Model and inference.}

We use Llama-3.2-1B-Instruct~\citep{grattafiori2024llama} as the retriever, fully
fine-tuned rather than adapted with a parameter-efficient method. The training data used
the examples from \S\ref{sec:graph-supervision}. At inference, we decode each facet
separately with multiple beams, constrained by the DocID trie and fuse with graph-RRF.
Exhaustive hyperparameters are in Appendix~\ref{app:hyper}.

\subsection{Baseline Comparison}
\label{sec:main-results}

Table~\ref{tab:xp-main} reports the main comparison. The full generative method,
\Method, reaches R@20 0.326, recovering 91\% of the dense-graph retriever (0.357). It
outperforms every baseline other than the graph.

\begin{table}[h]
\centering
\resizebox{\columnwidth}{!}{%
\begin{tabular}{lcc}
    \toprule
    \textbf{Method} & \textbf{R@20} & \textbf{R@50} \\
    \midrule
    Dense-graph & \textbf{0.357} & \textbf{0.560} \\
    MiniLM (title+abstract) & 0.295 & 0.440 \\
    Dense-only (facet) & 0.274 & 0.440 \\
    BM25 (title+facets) & 0.274 & 0.433 \\
    SPECTER2 (title+abstract) & 0.220 & 0.341 \\
    S2 Recommendations API & 0.124 & 0.129 \\
    MINDER & 0.284 & 0.392 \\
    \midrule
    Generative, uniform sampling & 0.296 & 0.430 \\
    \quad + edge-weighted oversampling & 0.307 & 0.442 \\
    \quad + coverage floor ($K_{\text{min}}{=}3$) & 0.322 & 0.447 \\
    \quad + \textbf{reverse-neighbour fallback (\Method)} & \textbf{0.326} & \textbf{0.453} \\
    \bottomrule
\end{tabular}
}
\caption{Main results. The full generative method approximates the dense-graph baseline
    at R@20 without a nearest-neighbour index or query-time encoder. The lower block is
    \Method's supervision comparison.}
\label{tab:xp-main}
\end{table}

The generative behaviour is worse at larger ranks. At R@50, the generative retriever
reaches 0.453, behind dense-graph (0.560). This suggests that the model learned the
top-rank behaviour of the graph pipeline, but deeper neighbourhoods are only partly
recovered. The gain from the reverse-neighbour fallback concentrates precisely there
($0.447 \rightarrow 0.453$), consistent with newly decodable tail papers entering the
lower ranks.

\subsection{Out-of-Corpus Queries}
\label{sec:out-of-corpus}

So far, we evaluated indexed papers, queries drawn from the same corpus the graph is
built on. \emph{Can the retrievers still find related works for unseen papers?} We
sample 500 papers from the larger pool of \S\ref{sec:corpus-construction}, so no query
is a corpus paper, and extract their facets with the same pipeline. The lexical and
document-level baselines run unchanged. The two graph-aware systems need adjustments:
the dense-graph retriever gets ephemeral edges between the new query and the corpus,
built with the same encoder and index, while \Method has no adjacency for graph-RRF, so
we estimate edge weights by mutual corroboration among the retrieved candidates.
Appendix~\ref{app:out-of-corpus} details the query set and both procedures.

\begin{table}[h]
\centering
\small
\begin{tabular}{lcccc}
    \toprule
    \textbf{Method} & \textbf{R@5} & \textbf{R@10} & \textbf{R@20} & \textbf{R@50} \\
    \midrule
    Dense-graph & 0.218 & 0.360 & 0.501 & 0.657 \\
    BM25        & 0.164 & 0.230 & 0.309 & 0.434 \\
    MiniLM      & 0.152 & 0.225 & 0.293 & 0.408 \\
    SPECTER2    & 0.121 & 0.170 & 0.225 & 0.314 \\
    \midrule
    \Method & \textbf{0.367} & \textbf{0.453} & \textbf{0.548} & \textbf{0.709} \\
    \bottomrule
\end{tabular}
\caption{Retrieval for 500 query papers outside the corpus. \Method leads at every
    cutoff. Baseline configurations are those of Table~\ref{tab:xp-main}.}
\label{tab:xp-ooc}
\end{table}

Table~\ref{tab:xp-ooc} shows the opposite behaviour of Table~\ref{tab:xp-main}.
In-corpus, \Method is bounded by the graph it distils, recovering 91\% of its R@20; on
unseen queries it leads at every cutoff (e.g., 0.501 to 0.548 R@20, $+9.4$\%) despite being the
simpler system. The other baselines underperform significantly, indicating the graph
signal remains essential. This suggests that when memorising the graph, the generative
retriever appears to develop a mapping from facets to related work that transfers to unseen
papers, whereas the dense-graph retriever relies on surface embedding similarity.

This transfer is driven by \emph{non-obvious retrievals}. We score every corpus paper
against the query using the same facet-item similarity used to build the graph
(\S\ref{sec:graph-teacher}), and call a retrieved paper \emph{obvious} if it is among
the query's 50 nearest neighbours and \emph{non-obvious} otherwise. Only 33.4\% of
\Method's retrievals are obvious, against 81.0\% of the dense-graph's, and 47.4\% of
\Method's correct retrievals are non-obvious; nearly half its recall lies beyond what
plain similarity reaches. These non-obvious picks are valid related papers: across the
graph, only 0.23\% of non-obvious candidates are cited by their query, compared with
7.1\% of those retrieved by \Method, a $30.8\times$ lift. This is not mere popularity;
against a baseline that always returns the most-cited non-obvious papers, \Method's
non-obvious retrievals have $4.2\times$ the precision. Appendix~\ref{app:qual-examples}
gives judged examples per facet and more details.

\subsection{Coverage} \label{sec:coverage-results}

\paragraph{Coverage-aware distillation.}

The lower block of Table~\ref{tab:xp-main} isolates each step of
\S\ref{sec:graph-supervision}. Every component improves recall. The two coverage
mechanisms together account for roughly two thirds of the gain, indicating that
\emph{which} DocIDs are trained matters as much as how the edges are weighted. Their
effect on coverage is direct: under uniform sampling only 84\% of the corpus ever
appears in the target position; the remaining papers are decodable through the trie but
the model never learns to generate them. The coverage floor and reverse-neighbour
fallback raise this coverage to the full 11,359 papers, confirming that part of the gain
comes from fixing DocID starvation.

\paragraph{The coverage floor is non-monotonic.}

We explore the coverage floor $K_\text{min}$ in Appendix~\ref{app:docid-schemes}. R@20
rises from 0.299 at one view per DocID to 0.322 at three, then falls to 0.294 at four.
This shows two issues: too few views leave a DocID unlearned and therefore undecodable,
while too many dilute the model with synthesised pairs that carry no additional graph
evidence, only noise. Three views is the peak.

\paragraph{Supervision alternatives.}

We also tried alternatives to the supervision design of \S\ref{sec:graph-supervision},
all at matched configuration on the uniform-sampling base (R@20 0.296), without the
coverage supervision. The closest, weighting the cross-entropy loss by edge score rather
than replicating pairs, reaches 0.303 but stays below cost-proportionate sampling
(0.307): loss weighting rescales the gradient on pairs already present, whereas
replication changes how often a DocID is seen. Appendix~\ref{app:supervision-alts} gives
the full comparison.

\subsection{Fusion and Attribution}
\label{sec:attribution}

\paragraph{Fusion.}

Table~\ref{tab:xp-rrf} compares ways of combining the four per-facet lists. A single
facet offers little on its own (R@20 0.115), as it is not enough to cover the full gold
set. Fusing all four with plain RRF reaches 0.284, and weighting that fusion by the
query--candidate graph using graph-RRF raises it to 0.322 (+0.038, +13.38\%). We also
compare against MINDER's combination, which ranks by language-model score rather than by
the graph and reaches R@20 0.310. It is ahead at R@50, but behind at R@20 and without
the facet attribution a graph weight carries. Appendix~\ref{app:fusion-ablations}
reports the per-facet ablation.

\begin{table}[h]
\centering
\small
\begin{tabular}{lcc}
    \toprule
    \textbf{Fusion} & \textbf{R@20} & \textbf{R@50} \\
    \midrule
    Best single facet (problems) & 0.115 & 0.115 \\
    Uniform RRF (no graph) & 0.284 & 0.447 \\
    MINDER combination & 0.310 & \textbf{0.464} \\
    Graph-RRF (all four facets) & \textbf{0.322} & 0.447 \\
    \bottomrule
\end{tabular}
\caption{Comparison between fusion methods.}
\label{tab:xp-rrf}
\end{table}
\vspace{-10pt}

\paragraph{Attribution transfers with the results.}

Recall measures \emph{which} papers the retriever returns, not \emph{why} it returns
them. \Method reproduces the graph's attribution at \textbf{0.922} precision, similarly
across the four facets (0.933 precision for results, 0.931 problems, 0.915
contributions, 0.903 methods). We measure this over every returned paper--query pair,
comparing the facets that graph-RRF credits against those the graph attributes to the
same connection, micro-averaged over those pairs. In 89.9\% of pairs, the credited set
is a subset of the reference: the model selects among facets the graph already supports
rather than inventing reasons. This shows that the generative retriever reproduces not
only which papers the graph returns but the provenance attached to them.

Appendix~\ref{app:attribution-example} works through a single query, where three facets
return topically related papers and the methods facet returns one related only by
methodology.

\subsection{Ablation studies}
\label{sec:further-ablations}

\paragraph{DocID representation.}

We compare three identifier schemes under uniform sampling, so that differences are
attributable to the identifier rather than coverage-aware distillation.
Natural-language facet-item DocIDs reach R@20 0.296, against 0.260 for cluster-keyword
(C2T) identifiers and only 0.121 for a numeric codebook, a $2.4\times$ gap.

The ordering follows how much pretrained language prior each scheme can reuse. Both
baselines create labels using hierarchical $k$-means clustering. The codebook uses
cluster indices, which carry no pretrained meaning. C2T uses keywords, which are natural
language the model understands but describe a \emph{cluster} rather than a single paper,
so papers under a leaf collide. The facet-item bullet is the paper's own text.
Appendix~\ref{app:docid-schemes} describes both baselines and
Appendix~\ref{app:identifier-construction} gives examples.

\paragraph{Capacity.}

Capacity is the main constraint on this task. Varying the base model, R@20 falls from
0.322 at 1B parameters to 0.248 at 0.6B and 0.089 at 360M, and low-rank adaptation
with the 1B model saturates at rank 256 (R@20 0.237), well short of full fine-tuning
(Appendix~\ref{app:capacity}). Reproducing verbatim bullet text is memorisation-bound,
and parameter-efficient tuning is not enough.

\section{Conclusion}
\label{sec:conclusion}

We posed two research questions: how to represent the heterogeneous relationships
between scientific papers, and how to index that structure efficiently for retrieval.
\Dataset answers the first with a typed paper graph whose edges carry both facet and
strength; \Method answers the second by distilling that graph into a generative
retriever, preserving both which papers are returned and the facets that explain each
connection. Two graph properties do not survive naive distillation: \emph{coverage} and
\emph{structural grounding}. We proposed \emph{coverage-aware distillation} and
\emph{graph-weighted RRF} to address each of these respectively. On our constructed
\Dataset, a corpus of 11,359 NLP papers, \Method approaches its graph teacher's recall
as a single model with only a decoding trie and a lightweight sparse adjacency matrix,
reproduces the graph's facet attribution, and outperforms the teacher on query papers
outside the corpus without needing the index. As future work, we aim to expand the
corpus beyond NLP, where a larger, sparser graph would test both graph-construction cost
and memorisation capacity.

\section*{Limitations}
\label{sec:limitations}

The corpus is narrow, covering only \textsc{*ACL} venues from 2019 to 2026, and our
$k$-core sampling favours well-connected papers, to the detriment of less prominent
work. Scaling is also a potential issue: graph construction is quadratic in the number
of nodes, and the retriever is limited by memorisation capacity. It is unknown what
model size is required as the graph grows.

Our comparison of deployment requirements covers the retrieval stage only. Both the
dense-graph retriever and \Method take the query paper's facet items as input, so a
paper that is not already in the corpus must have its facets extracted first, as was
done for the out-of-corpus queries of \S\ref{sec:out-of-corpus}. Moving the index
into the model weights also makes the corpus harder to update: a new paper can be
queried without retraining, as \S\ref{sec:out-of-corpus} shows, but \Method must be
retrained before it can return new papers. This is a known limitation of generative
retrieval~\citep{dsipp-mehta-2023}, which we do not address. This suits corpora that
grow in batches rather than continuously, which is the case for S2ORC, which is released
in periodic snapshots.

\section*{Ethics Statement} \label{sec:ethics}

The corpus is derived from open-access papers in S2ORC and involves no human subjects;
the only annotation is the facet-groundedness study of Appendix~\ref{app:groundedness},
carried out by three of the authors. The computational footprint is modest: facet
extraction cost approximately \$22 in API calls and the retriever is a single fine-tune
of a 1B-parameter model. Our sampling method favours well-connected papers by
construction, reinforcing existing biases in scientific work. Finally, our facets are
model-generated interpretations of actual papers, and the groundedness study covers only
a sample of them; inaccurate claims can come from our method rather than the original
work.

\bibliography{references}

\clearpage
\appendix

\section{Dataset Construction Details}
\label{app:td-corpus}

\subsection{Venue and Year Selection}

We build the corpus from S2ORC \citep{s2orc-lo-2020}, accessed through the Semantic
Scholar Datasets
API~\citep{kinney-2023-s2-platform},\footnote{\url{https://api.semanticscholar.org/api-docs/datasets}}
using release \texttt{2026-03-10} of the \texttt{s2orc\_v2} dataset, which supplies
paper metadata, citation edges, and parsed full text. We select papers by three
criteria:

\begin{itemize}

  \item \textbf{Venue.} We match the publication-venue string against an allowlist of
      the main \textsc{*ACL} and affiliated NLP venues: ACL, EMNLP, NAACL, COLING, EACL,
        AACL(-IJCNLP), CoNLL, SemEval, {*}SEM, TACL, and Computational Linguistics (CL).

  \item \textbf{Year.} Papers published 2019--2026, bounded by the 2026-03-10 release.

  \item \textbf{Full text.} We require a parsed full-text body, since our facets are
      extracted from the paper body. Metadata-only records are discarded.

\end{itemize}

\subsection{Citation-Preserving Sampling}

Facet extraction, embedding, and dense evaluation make the full filtered pool costly, so
we downsample to a tractable subset of 10,000. Naive uniform sampling does not work
well: a citation edge survives only when both endpoints are selected. Sampling $p$ nodes
keeps only ${\approx}\,p^{2}$ of the edges, likely resulting in a very sparse graph.
Instead, we use an alternative method to preserve citation density.

First, we take a $k$-core of the graph~\citep{seidman1983kcore} ($k = 3$) from
2019--2024. This means taking the maximal subgraph where each node (paper) is linked to
at least three other nodes also in the subgraph. This gives a well-connected starting
point. This 3-core contains 23,365 papers, more than twice our target of
10,000, so we select a subset from it.

Ranking papers purely by their number of connections favours older papers, which have
had more time to accumulate citations, to the detriment of recent work. To prevent this,
we assign each year a quota proportional to its share of the full corpus, using the
largest-remainder method so the quotas sum to exactly 10,000. Within each year, we
keep the highest-degree papers up to that year's quota.

A year whose pool cannot meet its quota is completed from the one-hop fringe neighbours
of the papers already selected. Candidates are ranked by how many edges they have into
the selected set, so the papers brought in are those most strongly attached to the
sample.

Finally, we extend the graph to 2025 and 2026 papers. This was done later, since these
papers are important for the overall quality of the corpus, but are poorly connected to
existing papers, and would likely be left behind by both procedures. The result is dense
and connected -- 112,644 citation edges at mean degree 22.5 -- rather than the
fragmented graph a random draw of the same size would give.

\subsection{Facet Extraction and Validation}\label{app:td-validation}

\paragraph{Extraction.} Each paper's facets come from a single request to
\texttt{gpt-5-mini-2025-08-07}. One prompt asks for all four facets, with one field per
facet: problems, then methods, results and contributions.

\begin{promptbox}[title=Facet extraction prompt]
System:
You are an expert at analysing academic papers. Extract the requested facets accurately and concisely. Always respond with valid JSON.

User:
Analyse this academic paper and extract the following facets.
For each facet, provide 1-5 distinct items (1-2 sentences each).

1. **Problems**: What problems or research questions does this paper address?
2. **Methods**: What approaches, methods, or techniques does the paper propose or use?
3. **Results**: What are the main results or findings?
4. **Contributions**: What are the contributions or novelties of this work?

Paper:
{paper_text}
\end{promptbox}

The schema enforces types only. It declares four fields, each a collection of strings,
and sets no item count restrictions and no length limit. The requirements are stated in
the prompt and are not enforced by the API.

Sampling is not under our control, so extraction is not fully reproducible. The
\texttt{gpt-5} family does not accept a temperature setting, and the structured-output
endpoint accepts no seed; we set a reasoning effort of \texttt{minimal}.

The prompt carries the whole paper: title, authors, and every parsed section under its
heading, head-truncated at 100,000 tokens with an explicit truncation marker.
Truncation is rare, as the median paper is 13,000 tokens and the 90th percentile is
69,000.

Returned strings are stored verbatim. A response that fails schema validation is
discarded and the paper left unprocessed. The audit figures describe the only attempt, as we do not retry papers that fail the audit.
The shipped corpus cost approximately \$22 in API charges. No local compute is used for
extraction.

\paragraph{Structural.} The prompt requests 1--5 items per facet of 1--2
sentences each. The audit uses more lenient parameters: every facet non-empty and at
most 5 items; every item between 10 and 500 characters and at most 3 sentences;
and no exact-string duplicates within a facet. We additionally detect meta-descriptions
(``The paper addresses\ldots'', ``The authors\ldots'' through five regular-expression
patterns) and penalise a paper when a whole facet or more than half of its items are
meta-descriptive. Violations are rare. Across the corpus we observe 33 over-length
items, 32 items exceeding the sentence limit, 16 fully meta-descriptive facets, 11
facets with too many items, one empty item and one exact duplicate; over 99\% of
papers have no issues at all.

\paragraph{Grounding.} We embed every facet item and every paragraph of its source
paper, and take an item's grounding score to be its maximum cosine similarity over all
paragraphs. This aims to verify that the extracted facet item is properly grounded in the paper content, not hallucinated. No section is selected in advance: the matching paragraph is simply the
argmax, and the three best-matching paragraphs are retained as evidence snippets. The
mean maximum cosine is 0.718, with per-facet means between 0.707 and 0.728.

\paragraph{Redundancy.} We compute cosine similarity between all facet-item embeddings
within a paper. Intra-facet pairs with similarity above 0.92 count as near-duplicates
and cross-facet pairs above 0.90 as overlaps. Both are uncommon: one paper in 11,359
contains an intra-facet near-duplicate, and 293 (2.9\%) contain at least one
cross-facet overlapping pair, usually between contributions and results.

\paragraph{Specificity.} We score how concrete each item's phrasing is with a lexical
heuristic, with the goal of avoiding generic statements. A meta-description prefix contributes 0.3 to a genericness score; each
match against 15 boilerplate patterns (``state-of-the-art'', ``extensive
experiments'') adds 0.2, capped at 0.5; and specific tokens (numbers, proper nouns,
acronyms, hyphenated technical terms) subtract 0.06 each, capped at 0.3. Specificity
is one minus this quantity, averaged over items. The corpus mean is 0.998 (p10
0.991).

\subsection{Facet Groundedness Study}
\label{app:groundedness}

The grounding check above is a proxy: a high cosine shows an item is close to some
paragraph, but it does not guarantee the paper supports its claim. To validate
groundedness directly, three annotators rated extracted facet items against their source
papers.

\paragraph{Sampling and protocol.}

We sample 50 papers whose extraction yielded all four facets, each one contributing one
item per facet (200 items). The annotation sheet shows the item alongside the paper's
title, abstract, introduction, and a link to the full PDF. Each item is rated 1--5 for
groundedness with a required one-sentence rationale: 5 is fully grounded, 4 minor
imprecision, 3 captures something real but a substantive part is unsupported, 2 only
loosely connected, and 1 fabricated or not about the paper. 40 items form a core rated
by all three annotators, on which we compute agreement. The remaining items are split
between annotators, extending coverage without duplicating effort. The three sheets
comprise 92, 92 and 96 items, for 280 ratings in total.

\paragraph{Ratings.}

No item is rated 1, and 3/280 ratings fall below 3. The annotators' rating distributions
(5/4/3/2) are 80/10/1/1, 63/23/6/0 and 89/3/2/2, with means of 4.84, 4.62 and 4.86. Each
annotator's own share of items rated ${\geq}\,4$ is 97.8\%, 93.5\% and 95.8\%, so all
three independently place groundedness above 93\%. Pooling over the 200 distinct items,
with each core item averaged across its three raters, 96.5\% are rated ${\geq}\,4$ (95\%
CI [93.0, 99.0]) with a mean of 4.78. Table~\ref{tab:app-groundedness} breaks the pooled
ratings down by facet: every facet's mean exceeds 4.6, with contributions the most
grounded facet and problems the least.

\begin{table}[t]
\centering
\small
\begin{tabular}{lcc}
    \toprule
    \textbf{Facet} & \textbf{Mean} & \textbf{${\geq}\,4$} \\
    \midrule
    Problems      & 4.68 & 47/50 (94\%) \\
    Methods       & 4.83 & 49/50 (98\%) \\
    Results       & 4.77 & 47/50 (94\%) \\
    Contributions & 4.85 & 50/50 (100\%) \\
    \midrule
    All           & 4.78 & 193/200 (96.5\%) \\
    \bottomrule
\end{tabular}
\caption{Pooled groundedness ratings by facet, with core items averaged across their
    three raters: mean score and fraction of items rated ${\geq}\,4$.}
\label{tab:app-groundedness}
\end{table}

\paragraph{Agreement.}

On the 40-item shared core, all three annotators assign an identical score to 27 items
(67.5\%) and unanimously rate 35 (87.5\%) as grounded (${\geq}\,4$); pairwise, exact
agreement is 70--82.5\% and agreement within one point 92.5--95\%. Chance-corrected
coefficients on the full scale are modest -- ordinal Krippendorff's
$\alpha$~\citep{krippendorff2018content} is 0.30 (95\% bootstrap CI [0.05, 0.53]) -- and
binarising at the ${\geq}\,4$ threshold does not improve them ($\alpha = 0.25$). This is
the expected behaviour of marginal-based coefficients under extreme class imbalance, not
evidence of unreliable annotation~\citep{feinstein1990high}. We therefore report Gwet's
$\mathrm{AC}_1$~\citep{gwet2008computing} for the grounded decision, which estimates
chance agreement without relying on the marginals and remains interpretable at this
prevalence: $\mathrm{AC}_1 = 0.91$ (95\% CI [0.81, 0.98]). Disagreements concern degree,
not groundedness. Most non-unanimous items differ only on the 4-vs-5 distinction. The
strictest annotator still rates 93.5\% of their items grounded.

\subsection{Dataset Statistics}

Table~\ref{tab:app-year-split} breaks the corpus down by publication year and split. The
splits are contiguous in time: training covers 2019--2022, development is 2023 alone,
and test spans 2024--2026.

Table~\ref{tab:app-venues} gives the twenty most frequent venues. Venue strings are
normalised to 246 distinct values, and every one of the 11,359 papers carries one.
Main conferences and journals account for 8,769 papers (77.2\%), the Findings
tracks for 656 (5.8\%), and workshops and other venues for the remaining 1,934
(17.0\%).

The four facet graphs are exact top-20 nearest-neighbour graphs over all 11,359
papers, so each contains 227,180 directed edges, with the four totalling 908,720
edges. This is the structure used during inference for graph-RRF, and it is considerably
smaller than the dense index it replaces.

\begin{table}[t]
\centering
\small
\begin{tabular}{lrrr}
    \toprule
    \textbf{Year} & \textbf{Train} & \textbf{Dev} & \textbf{Test} \\
    \midrule
    2019      & 1,479 &       &       \\
    2020      & 1,727 &       &       \\
    2021      & 1,839 &       &       \\
    2022      & 1,830 &       &       \\
    2023      &       & 1,543 &       \\
    2024      &       &       & 1,582 \\
    2025--26  &       &       & 1,359 \\
    \midrule
    Total     & 6,875 & 1,543 & 2,941 \\
    \bottomrule
\end{tabular}
\caption{Corpus by publication year and split.}
\label{tab:app-year-split}
\end{table}

\begin{table}[t]
\centering
\small
\begin{tabular}{r@{\hspace{6pt}}l r@{\hspace{6pt}}l}
    \toprule
    \textbf{$n$} & \textbf{Venue} & \textbf{$n$} & \textbf{Venue} \\
    \midrule
    2,891 & ACL               & 137 & TACL \\
    2,842 & EMNLP             & 135 & AACL-IJCNLP \\
    1,273 & NAACL             & 111 & CoNLL \\
    438   & EACL              & 102 & WMT \\
    398   & LREC              & 91  & INLG \\
    335   & Findings of ACL   & 78  & SIGDIAL \\
    256   & COLING            & 73  & BlackboxNLP \\
    180   & Findings of EMNLP & 72  & IWSLT \\
    168   & SemEval           & 69  & IJCNLP-AACL \\
    139   & Findings of EACL  & 51  & {*}SEM \\
    \bottomrule
\end{tabular}
\caption{The twenty most frequent venues in the corpus, after venue-string
    normalisation.}
\label{tab:app-venues}
\end{table}

\section{Graph Teacher Details}
\label{app:graph-teacher}

\subsection{Edge Construction and Weight Selection}
\label{app:edge-weight-selection}

The facet score between papers $A$ and $B$ under facet $f$ is the maximum-similarity
score of \S\ref{sec:graph-teacher}, averaged over each paper's items and over both
matching directions, where $a_i$ and $b_j$ are the embedded facet items of $A$ and $B$:

\begin{equation}
\begin{split}
    s_f(A, B) = \frac{1}{2}\Big(
        &\frac{1}{|A|} \sum_i \max_j \cos(a_i, b_j) \\
        &+ \frac{1}{|B|} \sum_j \max_i \cos(a_i, b_j)
    \Big).
\end{split}
\label{eqn:maxsim}
\end{equation}

The score is then rescaled to $[0,1]$. The citation score of
\S\ref{sec:facets-and-signals} compresses the bibliographic-coupling count
$x_{\mathrm{bc}}$ and the co-citation count $x_{\mathrm{cc}}$, both heavy-tailed, and
averages them:

\begin{equation}
    c(A,B) = \frac{1}{2}
        \sum_{x \in \{x_{\mathrm{bc}},\, x_{\mathrm{cc}}\}}
        \frac{\log(1+x)}{\log(1+x_{\max})},
    \label{eqn:citation-score}
\end{equation}

where $x_{\max}$ is the largest value of the corresponding count over the corpus. Edge
weights combine the two scores:

\begin{equation}
    w_f(A,B) = \alpha s_f(A,B) + \beta c(A,B).
    \label{eqn:facet-combination}
\end{equation}

We select
$\alpha$ by sweeping it over $\{0, 0.25, 0.5, 0.75, 1\}$ with $\beta = 1 - \alpha$,
evaluating the dense-graph retriever on the \textbf{development} split (1,543 papers,
2023). Table~\ref{tab:app-alpha} reports the sweep.

$\alpha = 0.75$ is the best setting on both metrics, and it is what we use in the main
method. The experiment also shows that neither signal is redundant. Used alone, the
citation score ($\beta = 1$, R@20 0.323) is the strongest, ahead of facet similarity
alone ($\alpha = 1$, R@20 0.273). However, any mixture of the two beats either in
isolation. The facet score carries most of the weight, while the citation score acts as
the floor described in \S\ref{sec:graph-teacher}, keeping an edge alive between papers
that co-occur in the literature even when their facet similarity is weak. This is especially important when creating the reverse neighbours.

\begin{table}[t]
\centering
\small
\begin{tabular}{cccc}
    \toprule
    \textbf{$\alpha$} & \textbf{$\beta$} & \textbf{R@20} & \textbf{R@50} \\
    \midrule
    0.00 & 1.00 & 0.323 & 0.524 \\
    0.25 & 0.75 & 0.339 & 0.552 \\
    0.50 & 0.50 & 0.356 & 0.568 \\
    0.75 & 0.25 & \textbf{0.363} & \textbf{0.569} \\
    1.00 & 0.00 & 0.273 & 0.437 \\
    \bottomrule
\end{tabular}
\caption{Edge-weight sweep over the facet/citation combination, dense-graph retrieval on
    the development split. $\alpha$ weights the facet score, $\beta = 1 - \alpha$ the
    citation score.}
\label{tab:app-alpha}
\end{table}

\subsection{Top-$K$ Pruning and Coverage}
\label{app:pruning}

The top-$K$ pruning of \S\ref{sec:graph-teacher} determines how much of the corpus is
reachable as a retrieval target: larger $K$ means more reachable candidates, but also
denser graphs and lower relatedness neighbours. Table~\ref{tab:app-pruning} shows how
changing $K$ affects coverage: at $K = 10$, only 52.9\% of papers have an incoming edge
in some facet, at $K = 20$ 84.0\%, and at $K = 50$ 98.5\%.

Raising $K$ more than doubles the number of gold pairs between $K = 20$ and $K = 50$,
enlarging both the graph and the adjacency matrix used for fusion. More importantly, a
paper's gold set is its neighbourhood in the graph (\S\ref{sec:experiments-setup}), so a
larger $K$ also includes weaker relations into the evaluation target, so coverage gained
through this might not translate to better results. The reverse-neighbour fallback of
\S\ref{sec:graph-supervision} instead uses the missing papers from edges already present
in the graph, giving full coverage at $K = 20$ without enlarging the neighbourhood.

\begin{table}[t]
\centering
\small
\begin{tabular}{lc}
    \toprule
    \textbf{$K$} & \textbf{Coverage} \\
    \midrule
    10 & 52.9\% \\
    20 & \textbf{84.0\%} \\
    50 & 98.5\% \\
    \bottomrule
\end{tabular}
\caption{Top-$K$ pruning level against target coverage. Our corpus uses $K = 20$
    (bold).}
\label{tab:app-pruning}
\end{table}

\subsection{Traversal Ablations}

The dense-graph retriever walks two hops from the query node
(\S\ref{sec:graph-teacher}), accumulating a path score multiplicatively along the
traversed edges and capping each facet at 500 candidates; the four facet lists are then
fused with RRF and truncated to the top 50. Table~\ref{tab:app-hops} compares walk
depths on the test queries. A single hop is clearly insufficient (R@20 0.318, R@50
0.465), since it restricts candidates to the query's own top-20 neighbours in each facet
and caps the reachable set well below the 39 gold neighbours an average query has.
Extending to two hops raises recall substantially, to 0.357 and 0.560. A third hop adds
nothing at R@20 and 0.005 at R@50, while enlarging the frontier and increasing the
computational cost of every traversal. We therefore use two hops in all experiments.

\begin{table}[t]
\centering
\small
\begin{tabular}{lcc}
    \toprule
    \textbf{Walk depth} & \textbf{R@20} & \textbf{R@50} \\
    \midrule
    1-hop & 0.318 & 0.465 \\
    2-hop (ours) & \textbf{0.357} & 0.560 \\
    3-hop & \textbf{0.357} & \textbf{0.565} \\
    \bottomrule
\end{tabular}
\caption{Walk depth for dense-graph retrieval on the test queries. Two hops capture
    almost all of the available recall; a third adds cost without benefit.}
\label{tab:app-hops}
\end{table}

\section{Generative Retriever Details}
\label{app:generative-details}

\subsection{Prompt and Target Format}
\label{app:prompt-format}

The retriever is queried once per facet item: a single item goes in and a single
identifier comes out. Prompts are rendered with the Llama-3 chat template with the
generation prompt appended. A single template serves all four facets, with the facet
name substituted; the four facets differ by that name alone.

\begin{promptbox}[title=Retrieval prompt]
System:
You are a generative retrieval system. Given a research paper's {Facet}, generate the semantic code of a relevant paper.

User:
{Facet}:
{item}
\end{promptbox}

Here \texttt{\{Facet\}} is one of \texttt{Problems}, \texttt{Methods}, \texttt{Results}
or \texttt{Contributions}, and \texttt{\{item\}} is a single facet item of the query
paper, entered as written in its original case and at full length.

The target is the identifier described in Appendix~\ref{app:identifier-construction}.

\subsection{Document Identifier Construction}
\label{app:identifier-construction}

\paragraph{Normalisation.} Each facet item becomes its own DocID: the unit is the item
rather than the facet, so a paper contributes roughly four identifiers per facet and
about sixteen in total. We normalise an item line by line. Leading bullet markers and
quotation characters are stripped, surrounding whitespace is removed, and the text is
lowercased; empty lines are dropped, and the surviving lines are joined with a single
space. Nothing further is applied: no Unicode normalisation, no punctuation stripping
beyond the leading markers, and no collapsing of interior whitespace. The identifier
therefore stays close to the paper's own wording.

\paragraph{Truncation.} Identifiers are truncated by token rather than by character to
12 tokens. This affects
 184,284 of the 184,288 (99.99\%) identifiers.

Table~\ref{tab:app-docid-example} shows the four identifiers a single paper contributes,
one per facet, beside the stored items they are drawn from. Truncation generally falls at
a phrase boundary rather than mid-word, so the identifiers remain readable English. What
it removes is trailing detail, such as a list of datasets,
architectures or metrics; the head of each claim, which carries its topical content,
is kept. This is why twelve tokens remain discriminative across the corpus, and why
identifiers of this kind can draw on the model's pretrained language prior in a way an
arbitrary code cannot.

\begin{table*}[t]
\centering\footnotesize
\begin{tabular}{@{}l p{0.45\linewidth} p{0.34\linewidth}@{}}
    \toprule
    \textbf{Facet} & \textbf{Stored facet item} & \textbf{Identifier (first 12 tokens)} \\
    \midrule
    Problems & Lack of systematic, comparable evaluation of post-hoc explainability
        techniques: existing studies focus on narrow setups (single method, task, or
        architecture) making it hard to choose suitable explainers for a given model and
        task. & lack of systematic, comparable evaluation of post-hoc explainability \\[4pt]
    Methods & Empirically test these diagnostic measures across three text-classification
        datasets (e-SNLI, TSE, IMDB) and three model architectures (CNN, LSTM,
        Transformer), including multiple random seeds and randomly initialized models to
        assess consistency. & empirically test these diagnostic measures across three
        text-classification datasets \\[4pt]
    Results & Rationale and dataset consistency measures show low-to-moderate
        correlations overall; no single explainer is universally best for rationale
        consistency, indicating that faithfulness, human-alignment, confidence
        indication, and consistency are separable properties. & rationale and dataset
        consistency measures show low-to-moderate correlations \\[4pt]
    Contributions & A comprehensive, practical suite of diagnostic properties for
        explainability methods together with explicit, automated measures (MAP, AUC-TP,
        SD+LR MAE, Spearman correlations) that are applicable across tasks and
        architectures. & a comprehensive, practical suite of diagnostic properties for
        explainability methods \\
    \bottomrule
\end{tabular}
\caption{The four identifiers contributed by \emph{``A Diagnostic Study of Explainability
    Techniques for Text Classification''}, one per facet, beside the stored facet items
    they derive from. Identifiers are lowercased, stripped of leading markers, and cut to
    twelve tokens.}
\label{tab:app-docid-example}
\end{table*}

\paragraph{Collisions.} Because identifiers are drawn from natural-language text rather
than assigned as codes, two papers can in principle produce the same one. Across the
corpus this is rare. Of 184,288 distinct identifiers, 41 (0.022\%) are shared by
more than one paper, at most three papers share any single identifier, and the mean
number of papers per identifier is 1.0003. They fall unevenly across facets, with 22
in problems against 6 in methods, 6 in results and 7 in contributions, consistent
with papers stating shared problems in similar language while describing their own
methods and findings in their own terms.

\paragraph{Tries.} We build one trie per facet, four in total, each a nested mapping
from token identifier to child nodes. Decoding runs once per facet for a given query,
and the four resulting lists are fused as described in \S\ref{sec:graph-rrf}. The tries
are built with the Llama-3 tokenizer, which matches the model used.
Table~\ref{tab:app-trie} gives their size: 1,794,934 internal nodes and
184,288 leaves in total, one leaf per distinct identifier, with every one of the
11,359 papers represented in all four facets. Dividing leaves by papers gives the
per-facet item counts of \S\ref{sec:facets-and-signals}: 3.70 for problems, 4.18 for
methods, 4.12 for results and 4.22 for contributions.

\begin{table}[t]
\centering
\small
\begin{tabular}{lrr}
    \toprule
    \textbf{Facet} & \textbf{Nodes} & \textbf{Leaves} \\
    \midrule
    Problems      & 396,133 & 41,997 \\
    Methods       & 481,796 & 47,490 \\
    Results       & 469,724 & 46,827 \\
    Contributions & 447,281 & 47,974 \\
    \midrule
    Total         & 1,794,934 & 184,288 \\
    \bottomrule
\end{tabular}
\caption{Per-facet prefix-trie size. Internal nodes exclude the root; each leaf is one
    distinct DocID.}
\label{tab:app-trie}
\end{table}

\subsection{Constrained-Decoding Algorithm}
\label{app:decoding}

Decoding is ordinary beam search with a generation constraint over the trie: given the
prefix generated so far, the trie determines which tokens can continue towards a valid
identifier, and the probabilities of all others are set to zero before the next token is
chosen. When a prefix reaches the end of every trie path, the end-of-sequence token is
forced to terminate the identifier. We generate with 50 beams and return all 50, without
sampling, using the end-of-sequence token for padding. Beams are ranked by the mean
log-probability per token.

\paragraph{Resolution.} Sequences are returned in descending score order.
A sequence that terminates anywhere other than a populated leaf is discarded. Where
an identifier maps to more than one paper, those papers enter the list in-corpus order
and take consecutive ranks.

\paragraph{Cost.} Every step evaluates the full 128,256-way output distribution, and
the constraint is applied by building a full-width mask that sets forbidden logits to
$-\infty$, costing $O(BV)$ per step irrespective of how few children the trie offers.
The trie lookups themselves are negligible beside the transformer forward passes.

\subsection{MINDER Baseline}
\label{app:minder-baseline}

The generative baseline of \S\ref{sec:main-results} replaces our identifier scheme and
our fusion while holding the rest of the setup fixed: the same base model, the same
training edges of \S\ref{sec:graph-supervision}, and the constrained decoding of
Appendix~\ref{app:decoding}.

Following MINDER~\citep{minder-li-2023}, a paper is represented by three kinds of
identifier: its title, contiguous substrings of its text, and synthetic pseudo-queries
generated from it. A paper carries many identifiers, as in \Method, but they are
untyped. There is no graph to weight a fusion with, so its lists are combined by
MINDER's own language-model score.

We build one trie per view type and decode each view separately, with the same beam
settings as Appendix~\ref{app:decoding}. Candidates are aggregated across views by the
language-model probability of the generated identifiers, and the paper's score is the
sum over the views that retrieved it. No other signal outside the model is used.

The full MINDER baseline reaches R@20 0.284 and R@50 0.392 (Table~\ref{tab:xp-main}).
Applying only MINDER's aggregation to \Method's four facet-item lists instead reaches
0.310 and 0.464 (Table~\ref{tab:xp-rrf}), so our facet-item identifiers contribute
+0.025 R@20 and +0.071 R@50 over MINDER's default views.

\section{Additional Results}
\label{app:additional-results}

\subsection{Out-of-Corpus Evaluation}
\label{app:out-of-corpus}

In \S\ref{sec:out-of-corpus}, we discussed evaluation results in an out-of-corpus
setting. In this setting, the evaluation queries come from outside the corpus entirely.
The papers in this subset have absolutely no connection to the 11,359 corpus at all. The
goal is to demonstrate how the dense-graph and generative retrievers perform when seeing
completely new papers, with the lexical and document-level dense baselines as reference
points. Those baselines run unmodified: they index the corpus and encode or match the
query's own text, so an unseen query is no different to them from an indexed one.

\paragraph{Query set.}

The filtered pool of Appendix~\ref{app:td-corpus} contains 27,757 qualifying papers, of
which $k$-core sampling retains 11,359. We draw 500 queries uniformly from the unused
papers, so the query is a valid paper that was not selected. Facets are extracted with
the same pipeline from Appendix~\ref{app:td-validation}. This means every paper is new
and was never seen during training.

\paragraph{Gold.}

Gold for a query is the set of papers it cites that are also in the new corpus, giving
2,095 query--paper pairs, a mean of 4.19 per query, median 3. We use direct citation as
the link between papers, as it was a connection type unused by the other experiments.
This ensures that there is no link between the main graph and our evaluation set.
Evaluation uses recall, micro-averaged over the 500 queries.

Absolute recall is higher than in Table~\ref{tab:xp-main} for every system. This is a
property of the citation dataset: a graph neighbourhood holds up to 20 neighbours per
facet, while a within-corpus bibliography holds 4.19 papers on average. This means the
retrieval results would naturally cover a large part of the gold papers.

\paragraph{Graph Fusion without Graph Edges.}

Graph-RRF (Equation~\ref{eq:graph-rrf}) scales each candidate's reciprocal-rank term by
the facet-graph edge between query and candidate. This is defined only when the query
$q$ is a corpus member, as the adjacency matrices include only indexed papers. This means
we cannot directly use graph-RRF on the out-of-corpus process since there is no edge
between the new papers and the indexed ones.

A possibility would be to encode the facets of the new papers to derive new edges for
the graph at query time. However, this would add the dense encoder and nearest-neighbour
index to the generative retrieval, which is against the very premise of the method.
Instead, we find a way to estimate the edge weight from the only available information:
the other candidates. We take the weight from mutual corroboration among the retrieved
candidates, which are corpus members and whose weights are always available. For a
retrieved set $\mathcal{C}$ and candidate $d \in \mathcal{C}$,

\begin{equation}
    \mathrm{corr}_f(d) = \sum_{p \in \mathcal{C} \setminus \{d\}}
        \max\big(s_f(d, p),\, s_f(p, d)\big),
    \label{eq:corr}
\end{equation}

\noindent and the fused score replaces $w_f$ with $\mathrm{corr}_f$:

\begin{equation}
    \mathrm{score}(d) = \sum_{f \in \facets \,:\, \mathrm{corr}_f(d) > 0}
        \frac{\mathrm{corr}_f(d)}{k + \mathrm{rank}_f(d)},
    \label{eq:corr-rrf}
\end{equation}

\noindent with $k = 60$ as before. The rationale is that papers genuinely related to the
query form a mutually connected neighbourhood in the facet graph. However, as seen in
the main graph-RRF, some decoding results will be completely unrelated to the query.
There, we exclude them based on the lack of an edge between query and decoded target.
Here, we exclude them when they are not corroborated by the other items, as an
approximation.

\paragraph{Ephemeral insertion for the dense-graph retriever.}

However, the dense-graph retriever has no comparable shortcut, as its walk needs to
start from a node. The solution is to insert the query ephemerally: its facet items are
encoded and scored against every item in the corpus with the same maximum-similarity
function used for edge construction (\S\ref{sec:graph-teacher}). The result is pruned to
the top 20 per facet, and the same two-hop walk is performed. This reproduces the
structure an indexed paper would have, so the baseline behaves the same as it would in
the main corpus. The cost is a full pass over the corpus index per query, which is what
\Method avoids.

\subsection{Non-Obvious Retrievals}
\label{app:qual-examples}

\S\ref{sec:out-of-corpus} shows that \Method's out-of-corpus retrievals concentrate in
non-obvious papers: those outside the query's 50 nearest under the graph's own
facet-similarity scoring. Of \Method's top-20 results, 66.6\% are non-obvious, against
19.0\% for the dense-graph retriever, and 47.4\% of \Method's correct retrievals come
from the non-obvious set: nearly half of its recall lies beyond what plain similarity
can retrieve.

\paragraph{Popularity control.}

Since \Method is trained on graph edges, it could achieve the lift of
\S\ref{sec:out-of-corpus} by defaulting to frequently cited papers: common influential
papers would be relevant to many queries, but would be non-specific and unhelpful. We
therefore compare against a baseline that always retrieves the most-cited non-obvious
papers. \Method's non-obvious retrievals have $4.2\times$ the precision of this
baseline, so citation frequency alone does not explain its behaviour.

\paragraph{Judged examples.}

Table~\ref{tab:app-qual-facets} gives one judged example per facet, each retrieved paper
ranked 74th to 92nd by similarity, immediately outside the 50-nearest threshold. In each
pair, the shared substance is distant enough enough to push the papers apart in the
similarity ranking, while the generative retriever is still able to recognise their
significance from its training.

\begin{table*}[t]
\centering
\small
\begin{tabular}{@{}lp{0.32\textwidth}cp{0.34\textwidth}@{}}
    \toprule
    \textbf{Facet} & \textbf{Query $\rightarrow$ retrieved paper} & \textbf{Rank} & \textbf{Shared, in different words} \\
    \midrule
    problems &
        Higher-order comparisons of sentence encoder representations $\rightarrow$
        Designing and interpreting probes with control tasks &
        92 &
        Whether an interpretability method measures the representation or the instrument:
        via representational geometry, and via probe expressivity. \\
    \addlinespace
    methods &
        Findings from the Bambara--French machine translation competition $\rightarrow$
        AmericasNLI &
        84 &
        Building a parallel resource for a genuinely low-resource language and adapting a
        multilingual pretrained model to it. \\
    \addlinespace
    results &
        Semantic representation for dialogue modeling $\rightarrow$
        Entity, relation and event extraction with contextualized span representations &
        74 &
        An explicit graph over a contextual encoder recovers long-range structure the
        encoder alone misses: across utterances, and across sentences. \\
    \addlinespace
    contributions &
        A challenge dataset for conversational stance detection $\rightarrow$
        Build it break it fix it for dialogue safety &
        80 &
        A multi-turn dialogue dataset built to capture what single-turn collection misses,
        for stance and for safety. \\
    \bottomrule
\end{tabular}
\caption{One judged non-obvious retrieval per facet. ``Rank'' is the retrieved paper's
    rank in the facet-similarity ranking.}
\label{tab:app-qual-facets}
\end{table*}

\subsection{Identifier Schemes and the Coverage Floor}
\label{app:docid-schemes}

Table~\ref{tab:app-docid} compares the three identifier schemes discussed in
\S\ref{sec:further-ablations}, all trained on the uniform-sampling base so that
identifier design is isolated from the training recipe. Table~\ref{tab:app-kmin} reports
the coverage-floor sweep over $K_{\text{min}}$.

Numeric and C2T baseline schemes share the same backbone: a hierarchical $k$-means
clustering of the facet-item embeddings, which assigns every paper a root-to-leaf path.
The \emph{numeric codebook} represents each paper as the cluster indices from that path,
so related papers share a prefix. However, these indices are still new vocabulary,
learnt from scratch. \emph{C2T}~\citep{zhang2025c2tid} keeps the same paths and replaces
each index with keywords extracted from the members of that cluster, so the code is
natural language instead of arbitrary tokens. The two schemes differ only in whether the
identifier is arbitrary or drawn from the model's vocabulary, which is the comparison
Table~\ref{tab:app-docid} isolates.

\begin{table}[t]
\centering
\small
\begin{tabular}{lc}
    \toprule
    \textbf{DocID scheme} & \textbf{R@20} \\
    \midrule
    Numeric codebook & 0.121 \\
    C2T cluster keywords & 0.260 \\
    Facet-item bullet & \textbf{0.296} \\
    \bottomrule
\end{tabular}
\caption{Identifier scheme (uniform-sampling baseline, graph-RRF R@20).}
\label{tab:app-docid}
\end{table}

\begin{table}[t]
\centering
\small
\begin{tabular}{cc}
    \toprule
    \textbf{$K_{\text{min}}$} & \textbf{R@20} \\
    \midrule
    1 & 0.299 \\
    2 & 0.304 \\
    3 & \textbf{0.322} \\
    4 & 0.294 \\
    \bottomrule
\end{tabular}
\caption{Coverage-floor sweep: minimum training views per DocID.}
\label{tab:app-kmin}
\end{table}

\subsection{Fusion Ablations}
\label{app:fusion-ablations}

Table~\ref{tab:app-loo} removes one facet at a time from the graph-RRF fusion of
\S\ref{sec:attribution}. Every facet is important and contributes to performance, most
for \emph{problems} (-0.020) and least for \emph{methods} (-0.002), so the problem facet
carries the most signal.

\begin{table}[t]
\centering
\small
\begin{tabular}{lcc}
    \toprule
    \textbf{Fusion} & \textbf{R@20} & \textbf{$\Delta$} \\
    \midrule
    Graph-RRF (all four facets) & \textbf{0.322} & \\
    \quad $-$ problems & 0.303 & -0.020 \\
    \quad $-$ methods & 0.321 & -0.002 \\
    \quad $-$ results & 0.317 & -0.005 \\
    \quad $-$ contributions & 0.316 & -0.006 \\
    \bottomrule
\end{tabular}
\caption{Graph-RRF with one facet removed ($K_{\text{min}}{=}3$ model), and $\Delta$ its
    change in R@20 from the full fusion.}
\label{tab:app-loo}
\end{table}

\subsection{A Facet Attribution Example}
\label{app:attribution-example}

Table~\ref{tab:xp-qual} shows what the per-facet decomposition provides. For the paper
\emph{``CEval: A Benchmark for Evaluating Counterfactual Text Generation''}, three
facets return counterfactual work, but the methods facet returns an LLM-as-a-judge paper
with no counterfactual content: relevant by methodology instead of topic, which
whole-paper similarity would never surface. Because the graph-RRF score decomposes into
per-facet terms (Eq.~\ref{eq:graph-rrf}), this provenance comes directly out of fusion.

\begin{table}[t]
\centering\footnotesize
\begin{tabular}{@{}l p{0.66\linewidth}@{}}
    \toprule
    \textbf{Query facet} & \textbf{Top retrieved related paper} \\
    \midrule
    Problems & Flexible Text Generation for Counterfactual Fairness Probing \\
    Methods & YESciEval: Robust LLM-as-a-Judge for Scientific Question Answering \\
    Results & Optimal and Efficient Text Counterfactuals Using Graph Neural Networks \\
    Contributions & DISCO: Distilling Counterfactuals with Large Language Models \\
    \bottomrule
\end{tabular}
\caption{Facet-attributed retrieval for the query paper \emph{``CEval: A Benchmark for
    Evaluating Counterfactual Text Generation''}. Each facet surfaces a different
    neighbour, retaining why it was recommended.}
\label{tab:xp-qual}
\end{table}

\subsection{Supervision and Objective Alternatives}
\label{app:supervision-alts}

Table~\ref{tab:app-supervision} reports the alternatives to our supervision design
discussed in \S\ref{sec:coverage-results}. All variants are trained on the
uniform-sampling base without the coverage floor or the reverse-neighbour fallback. The
upper block varies how the edge score is used, and is discussed in
\S\ref{sec:coverage-results}. The lower block covers two alternatives that instead
change the training data itself, and both fall well below the uniform base.

Enumerating every source--target bullet pair, rather than sampling one target item per
example, drops R@20 to 0.242. The exhaustive pairing floods training with combinations
that carry no additional graph evidence, and the resulting example distribution no
longer reflects edge importance.

Adding hard negatives with an unlikelihood term is even worse: 0.211 with a top-50
negative pool and 0.201 at top-100, costing more than any other choice in this
table. This suggests that negative examples are still similar enough to the positive
ones that this method only confuses training instead of producing better separation
between positive and negative entries.

\begin{table}[t]
\centering
\small
\begin{tabular}{lcc}
    \toprule
    \textbf{Variant} & \textbf{R@20} & \textbf{R@50} \\
    \midrule
    Uniform sampling (base) & 0.296 & 0.430 \\
    CE reweighting by edge score & 0.303 & 0.436 \\
    Graph-weighted sampling (ours) & \textbf{0.307} & \textbf{0.442} \\
    \midrule
    Exhaustive Cartesian pairing & 0.242 & 0.328 \\
    Hard negatives, top-50 & 0.211 & 0.277 \\
    Hard negatives, top-100 & 0.201 & 0.275 \\
    \bottomrule
\end{tabular}
\caption{Supervision and objective alternatives. Top: ways of using the edge score.
    Bottom: alternative training-pair construction and negative sampling.}
\label{tab:app-supervision}
\end{table}

\section{Model Capacity}
\label{app:capacity}

Table~\ref{tab:app-lora} compares LoRA ranks against full fine-tuning using the main
method and Llama-3.2-1B-Instruct model. LoRA at first improves with rank, but then
saturates: rank 128 achieves R@20 0.230, but rank 256 reaches only R@20 0.237, a
difference of merely +0.007. It performs well below full fine-tuning at every rank.
Together with the model-size differences of Table~\ref{tab:app-size}, it explains why we
fully fine-tune: the task is bound by memorisation capacity, and low-rank adaptors do
not provide enough of it. LoRA uses learning rate $2\mathrm{e}{-4}$ (versus
$5\mathrm{e}{-5}$ for full fine-tuning), otherwise identical hyperparameters.

\begin{table}[t]
\centering
\small
\begin{tabular}{lcc}
    \toprule
    \textbf{Model} & \textbf{R@20} & \textbf{R@50} \\
    \midrule
    LoRA $r{=}16$ & 0.169 & 0.232 \\
    LoRA $r{=}32$ & 0.205 & 0.282 \\
    LoRA $r{=}64$ & 0.217 & 0.293 \\
    LoRA $r{=}128$ & 0.230 & 0.314 \\
    LoRA $r{=}256$ & 0.237 & 0.337 \\
    \midrule
    Full fine-tuning & \textbf{0.322} & \textbf{0.447} \\
    \bottomrule
\end{tabular}
\caption{LoRA rank sweep vs.\ full fine-tuning, graph-RRF recall.}
\label{tab:app-lora}
\end{table}

Table~\ref{tab:app-size} varies the base model instead, holding the method fixed. Recall
degrades sharply as capacity falls, and the two smallest models fail to memorise the
corpus at all.

\begin{table}[t]
\centering
\small
\begin{tabular}{lcc}
    \toprule
    \textbf{Base model} & \textbf{R@20} & \textbf{R@50} \\
    \midrule
    Llama-3.2 1B & \textbf{0.322} & \textbf{0.447} \\
    Qwen3 0.6B & 0.248 & 0.332 \\
    SmolLM2 360M & 0.089 & 0.130 \\
    SmolLM2 135M & 0.084 & 0.112 \\
    \bottomrule
\end{tabular}
\caption{Base-model size (graph-RRF recall).}
\label{tab:app-size}
\end{table}

\section{Hyperparameters}
\label{app:hyper}

\paragraph{Retriever training.}

Base model Llama-3.2-1B-Instruct, fully fine-tuned (no LoRA); 2 epochs; batch size
4; gradient accumulation 4 (effective batch 16); learning rate $5\mathrm{e}{-5}$;
maximum sequence length 512 tokens (including prompt, DocID, and EOS).

\paragraph{Training-data construction (\S\ref{sec:graph-supervision}).}

Edge-weighted oversampling with replication ceiling $C=5$; coverage floor
$K_{\text{min}}=3$; reverse-neighbour fallback breadth $N=10$. Each example pairs a
single source bullet with one target item, which is sampled per example; every source
bullet of the query yields an example, so source items are never subsampled. Edge
weights are the per-facet candidate weights.

\paragraph{Inference.}

Beam width 50 per item under trie-constrained decoding. Each of the query's facet
items is decoded separately, and a facet's candidate list is the concatenation of its
items' results in bullet order, so a facet with four bullets contributes up to 200
candidates before fusion. Graph-RRF fusion with $k=60$ over the four facets.

\paragraph{Graph construction (\S\ref{sec:facets-and-signals}).}

Embeddings from \texttt{all-MiniLM-L12-v2}; edge-weight fusion via linear combination of facet
scores ($\alpha=0.75$) and citation scores ($\beta=0.25$); 100 candidates per paper
per facet pruned to the top 20 edges; two-hop BFS traversal with 10 diversity slots
and RRF $k=60$.

\paragraph{Packages used.} Table~\ref{tab:app-packages} shows the main packages used
during development with their versions and licences. A complete listing of packages with
full version specification is available with the source code.

\begin{table}[t]
\centering
\small
\begin{tabular}{lrc}
    \toprule
    \textbf{Package} & \textbf{Version} & \textbf{Licence} \\
    \midrule
    \texttt{accelerate}            & 1.12.0   & Apache-2.0 \\
    \texttt{beartype}              & 0.22.9   & MIT \\
    \texttt{bitsandbytes}          & 0.49.2   & MIT \\
    \texttt{fastapi}               & 0.135.1  & MIT \\
    \texttt{leidenalg}             & 0.12.0   & GPL-3.0 \\
    \texttt{matplotlib}            & 3.10.7   & PSF \\
    \texttt{networkx}              & 3.6.1    & BSD \\
    \texttt{numpy}                 & 2.2.6    & BSD \\
    \texttt{openai}                & 2.16.0   & Apache-2.0 \\
    \texttt{peft}                  & 0.19.1   & Apache-2.0 \\
    \texttt{pydantic}              & 2.12.5   & MIT \\
    \texttt{pymupdf4llm}           & 1.27.2.2 & AGPL-3.0 \\
    \texttt{python-igraph}         & 1.0.0    & GPL-2.0 \\
    \texttt{rank-bm25}             & 0.2.2    & Apache-2.0 \\
    \texttt{rapidfuzz}             & 3.14.5   & MIT \\
    \texttt{scikit-learn}          & 1.8.0    & BSD \\
    \texttt{sentence-transformers} & 5.3.0    & Apache-2.0 \\
    \texttt{socksio}               & 1.0.0    & ISC \\
    \texttt{tiktoken}              & 0.12.0   & MIT \\
    \texttt{torch}                 & 2.10.0   & BSD \\
    \texttt{transformers}          & 5.11.0   & Apache-2.0 \\
    \texttt{uvicorn}               & 0.42.0   & BSD \\
    \bottomrule
\end{tabular}
\caption{Python packages used during development.}
\label{tab:app-packages}
\end{table}

\section{Use of AI Assistants}
\label{app:ai-assistants}

LLM-based assistants were used for coding and literature search. All generated code was
reviewed and tested by the authors, and all papers obtained through automated literature
search were verified before being cited.

\end{document}